\documentclass{aa}
\usepackage{graphicx}
\usepackage{txfonts}
\begin{document}
   \title{Gas near active galactic nuclei: A search for the 
    4.7$\mu$m CO band
   \thanks
   {Based on observations with ISO, an ESA project with instruments funded
   by ESA member states (especially the PI countries: France, Germany, the
   Netherlands, and the United Kingdom) with the participation of ISAS and
   NASA.}
   }

   \subtitle{}
   \titlerunning{Search for the 4.7$\mu$m CO band in AGN}
   \authorrunning{Lutz et al.}

   \author{D. Lutz\inst{1}, E. Sturm\inst{1}, R. Genzel\inst{1},
          H.W.W. Spoon\inst{2} \and G.J. Stacey\inst{2}
          }

   \offprints{D. Lutz}

   \institute{Max-Planck-Institut f\"ur extraterrestrische Physik, Postfach 
              1312, 85741 Garching, Germany
              \\
              \email{lutz@mpe.mpg.de, sturm@mpe.mpg.de, genzel@mpe.mpg.de}
         \and
             Cornell University, Astronomy Department, Ithaca, NY 14853\\
             \email{spoon@isc.astro.cornell.edu, stacey@astro.cornell.edu}
             }

   \date{Received; accepted}

   \abstract{In order to constrain the properties of dense and warm gas
   around active galactic nuclei, we  have searched Infrared Space Observatory
   spectra of local active galactic nuclei (AGN) for the signature
   of the 4.7$\mu$m fundamental ro-vibrational band of carbon monoxide. 
   Low resolution spectra of 31 AGN put upper limits on the presence of 
   wide absorption bands corresponding to absorption by large columns of 
   warm and dense
   gas against the nuclear dust continuum. High resolution (R$\sim$2500) 
   spectra of
   \object{NGC\,1068} detect no significant absorption or emission
   in individual lines, to a 3$\sigma$ limit of 7\% of the continuum. 
   The limits 
   set on CO absorption in local AGN are much lower than the recent 
   Spitzer Space Telescope detection
   of strong CO absorption by dense and warm gas in the obscured ultraluminous
   infrared galaxy IRAS F00183-7111, 
   despite evidence for dense material on parsec scales near an AGN
   in both types of objects. 
   This suggests that such deep absorptions are not intimately related to the 
   obscuring `torus' material invoked in local AGN, but rather are a signature
   of the peculiar conditions in the circumnuclear region of highly obscured
   infrared galaxies like IRAS F00183-7111. They may reflect fully covered 
   rather than torus geometries.
    
   }

   \maketitle
%

\section{Introduction}
The circumnuclear regions of active galaxies, exposed to 
the intense nuclear UV and X-ray radiation field, host dense material in the
form of the clumpy `torus' posited by unified models and/or of more irregularly
distributed clouds. Under these conditions, significant quantities of warm
($\approx$1000K) and dense ($\approx 10^7$cm$^{-3}$)
molecular gas can be created (e.g. Krolik \& Lepp \cite{krolik89},
Maloney et al. \cite{maloney96}). While such gas can emit significantly
in familiar tracers of molecular material like the mm CO rotational lines or
the near-infrared H$_2$ rovibrational lines, these lines are also emitted
by molecular gas under less extreme conditions.
Focussing on transitions that 
originate only in warm/dense gas can help breaking these degeneracies, in 
particular if found at wavelengths able to penetrate significant
obscuring dust columns. 

\begin{figure}
\includegraphics[width=\columnwidth]{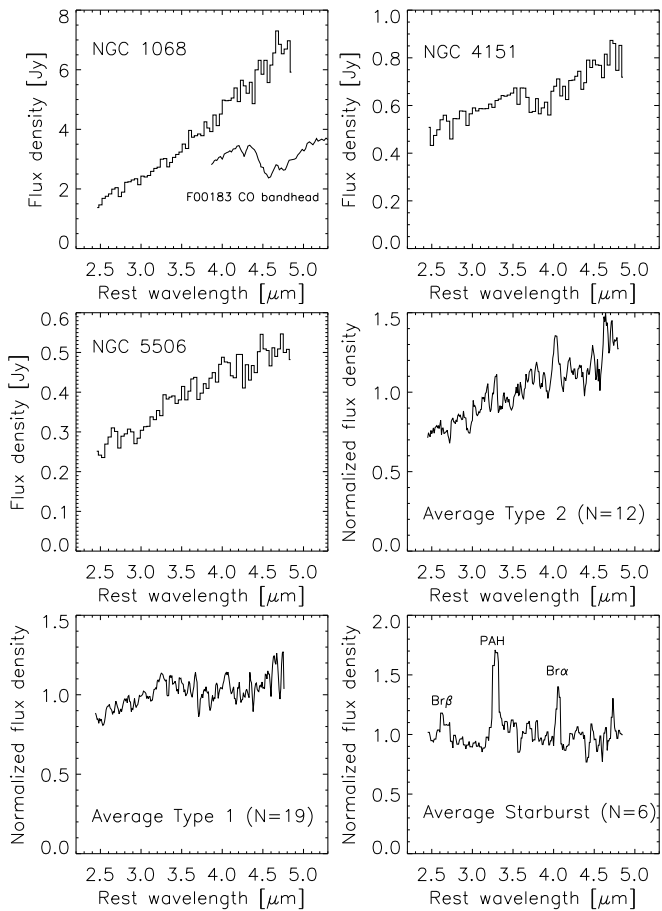}
\caption{ISOPHOT-S low resolution 2.5-5$\mu$m spectra of AGN. We show
three examples of good S/N spectra, plus averages for Type 1 and 
Type 2 Seyferts. These averages include fainter spectra, all normalized 
to median flux 1 in the observed range before averaging. They give 
equal weight to all objects irrespective of flux and S/N. A starburst 
spectrum is shown for
comparison and to identify features also seen in some of the AGN spectra.
In the top left panel, a scaled spectrum of IRAS F00183-7111
(Spoon et al. \cite{spoon04}) is included for comparison. 
}
\label{fig:phts}
\end{figure}

One of the possible tracers are highly excited CO transitions, either
in far-infrared rotational lines as modelled by Krolik \& Lepp 
(\cite{krolik89}), or using high rotational quantum number J transitions of 
the 1--0 rovibrational band near 4.7$\mu$m.
The lowest few transitions of the P and R branch of this band, superposed
on broad absorption features of CO and `XCN' ice, have been detected towards
galactic nuclei obscured by large columns of cold molecular material, like our
Galactic Center (Lutz et al. \cite{lutz96}, Moneti et al. \cite{moneti01})
or the circumnuclear starburst of \object{NGC\,4945} (Spoon et al. 
\cite{spoon03}). Rovibrational transitions from higher (J$\gtrsim$10) levels
are expected only from warm and dense regions, and have been observed
in emission (e.g. Scoville et al. \cite{scoville83}, Rosenthal et al.
\cite{rosenthal00}, Pontoppidan et al. \cite{pontoppidan02}) or 
absorption (e.g. Mitchell et al. \cite{mitchell89}) near young stellar 
objects in our galaxy. Similarly,
detection of high J rovibrational CO transitions near AGN would be direct
evidence for the presence of warm and dense gas, with the caveat that
complex geometries and radiative transfer can lead to difficult 
interpretation due to a complex interplay of line emission, line absorption and
continuous absorption. As expected from `torus' dust models and mm-wave CO 
observations and confirmed directly by Br$\alpha$ spectroscopy 
of Seyfert 2 galaxies (Lutz et al. \cite{lutz02}), the dust optical 
depth in the nuclear regions can still be considerable at 4--5$\mu$m. Part of
the warm dense gas can thus be hidden at those wavelengths. In addition, 
the observable signature
in the CO lines will depend on the temperature of regions that are optically
thin in the continuum compared to the temperature of the dust continuum 
`photosphere'. This can cause shallow line absorptions/emissions even in 
case of large columns of warm molecular gas.

The topic has recently gained further interest due to the Spitzer 
Space Telescope detection (Spoon et al. \cite{spoon04}) of
gaseous CO absorption indicating a large column of dense ($\rm\sim10^6cm^{-3}$)
and warm ($\sim$700K) gas
towards the nuclear region of the unusually obscured ultraluminous infrared
galaxy (ULIRG) IRAS F00183-7111,
most likely hosting an AGN. This detection has motivated us to use archival
data from the Infrared Space Observatory ISO 
to detect or put limits on absorption or emission from the
CO fundamental rovibrational band in bright local AGN. We
analyse low resolution ISOPHOT-S spectra of 31 nearby AGN allowing us to
put limits on the presence of strong and broad absorption bands of the type
observed in IRAS F00183-7111. For the prototypical Seyfert 2 NGC\,1068, a
high resolution and high S/N SWS spectrum allows us to put limits on the 
presence of individual CO transitions. We discuss the implications of these
nondetections for the conditions around nearby AGN, and for the interpretation
of strongly different spectra like that of IRAS F00183-7111.

\section{Limits from low resolution ISOPHOT spectra}
We have searched the ISOPHOT-S low resolution spectra of nearby 
(cz$<$10000km/s) AGN
for the presence of broad absorptions similar to the ones observed in IRAS 
F00183-7111. These observations will not resolve individual transitions, but
are sensitive to strong bands reaching up to high J values.
At these modest redshifts, much of the CO band is still included
in the short wavelength segment of ISOPHOT-S which extends to 4.9$\mu$m 
observed wavelength. We use the database of extragalactic ISOPHOT-S
spectra already exploited by Spoon et al. (\cite{spoon02}) and Lutz et al.
(\cite{lutz04}).
For signal-to-noise reasons, we have limited ourselves
to 31 bright AGN having a 6$\mu$m continuum (derived using the spectral
decomposition of Lutz et al. \cite{lutz04}) in excess of 0.1 Jy. For objects
with S/N good enough to individually exclude a feature of the type seen
in IRAS F00183-7111, we quote in 
Table~\ref{tab:limits} 3$\sigma$ limits for the depth of a gaussian 
of FWHM 0.4$\mu$m, which is a reasonable approximation to the IRAS F00183-7111
feature over the rest wavelength range covered. 

\begin{table}
\begin{tabular}{llr}\hline
Object          &Type         &Depth\\ \hline
IRAS F00183-7111&Obscured AGN?&0.40 \\ \hline
Ark 120         &Type 1    &$<$0.38\\
Cen A           &Type 2    &$<$0.10\\
Circinus        &Type 2    &$<$0.40\\
IC4329A         &Type 1    &$<$0.21\\
MCG8-11-11      &Type 1    &$<$0.25\\
Mrk 6           &Type 1    &$<$0.37\\
NGC 1068        &Type 2    &$<$0.12\\
NGC 1365        &Type 2    &$<$0.20\\
NGC 3516        &Type 1    &$<$0.33\\
NGC 3783        &Type 1    &$<$0.40\\
NGC 4151        &Type 1    &$<$0.12\\
NGC 5506        &Type 2    &$<$0.10\\
NGC 7213        &Type 1    &$<$0.16\\
NGC 7469        &Type 1    &$<$0.29\\
NGC 7582        &Type 2    &$<$0.37\\ \hline
Average Spectrum&Type 1    &$<$0.15\\
Average Spectrum&Type 2    &$<$0.09\\ \hline
\end{tabular}
\caption{Limits (3$\sigma$) on the depth of a 4.7$\mu$m CO rovibrational 
band of the type observed in IRAS F00183-7111, for some bright nearby 
AGN. Depth is expressed in fraction of the continuum.}
\label{tab:limits}
\end{table}

Fig.~\ref{fig:phts} illustrates this by three examples of spectra with 
good S/N. 
In addition we show average spectra of all 19 Type 1 and of all 12 Type 2
spectra meeting our basic redshift and brightness criteria. The average
spectra give equal weight to all objects irrespective of their brightness
and S/N. This is appropriate since source-to-source variations in 
feature properties are conceivable, making weighting towards the few 'best'
spectra undesirable. Like the 
individual brightest sources, the average spectra clearly do not show 
strong band 
absorption of the type seen in IRAS F00183-7111. The peaks seen around 
4.7$\mu$m in the average spectra are not significant emissions, given that the
noise at the long wavelength end of these ISOPHOT-S spectra is a factor $\sim$5
higher than near its minimum around 3$\mu$m. Br$\alpha$ 4.05$\mu$m and 
tentative 3.2$\mu$m PAH features are detected in the average spectra.

The nondetection of strong CO absorption applies both to Type 1 and to 
Type 2 objects. We have tested
this separately, since the latter are more likely to be absorbed from a 
point of view of AGN unification. The nondetection remains when 
taking the Type 2 average only for
9 objects with X-ray absorbing column $\rm N_H>10^{23}cm^{-2}$ - the resulting
average Type 2 spectrum is very similar to that shown. Since the CO 
absorptions searched for are sensitive only to absorbing material in a 
certain physical state and
not directly proportional to total column, and since X-ray
and infrared absorptions are towards different emitting regions of the AGN, 
the absence of strong CO absorptions towards Compton-thick AGN does not 
necessarily indicate a problem with the unified view.

\section{Limits from resolution $\sim$2500 SWS spectrum of NGC\,1068}
\begin{figure}
\includegraphics[width=\columnwidth]{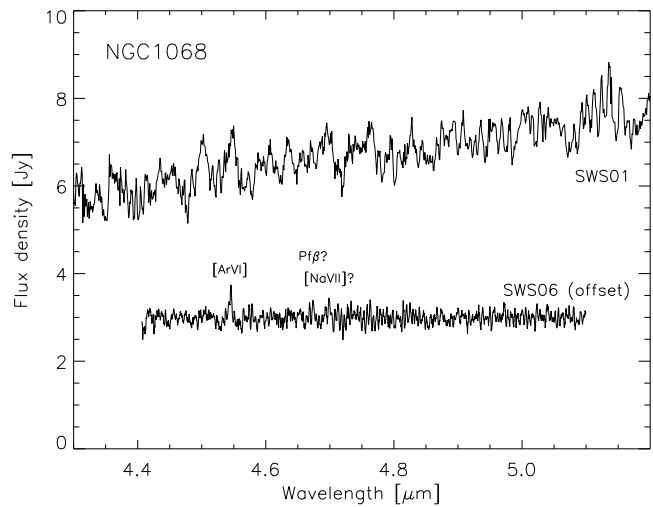}
\caption{The region of the CO fundamental rovibrational band in ISO-SWS 
spectra of NGC\,1068. Top: part of the full SWS01 spectrum also presented
by Lutz et al. (\cite{lutz00}). Bottom: SWS06 spectrum specially processed as
described in the text.
This specturm has better signal-to-noise and resolution $\sim$2500. Due to
the special processing, information on continuum slope and broad features is
lost, however, and the spectrum is shown at an arbitrary flux offset.}
\label{fig:sws}
\end{figure}

Since it is close, very bright in the mid-infrared continuum, and has a 
Compton-thick Seyfert 2 nucleus, NGC\,1068 is well suited to probe for 
individual
CO absorption lines. Lutz et al. (\cite{lutz00}) have presented ISO-SWS 
spectroscopy of NGC\,1068. The upper spectrum in Fig.~\ref{fig:sws} shows the
region around the CO band from the SWS01 full spectrum presented in that 
paper, confirming and strengthening by an independent dataset the 
absence of a strong and wide
CO absorption concluded in Sect.~2 from the ISOPHOT-S spectrum.
We estimate an upper limit of $\sim$10\% for the depth of any possible
broad feature.

A dedicated ISO-SWS observation (Observing mode SWS06, TDT number 79201901, 
see Leech et al. (\cite{leech03}) for details on SWS and its observing
modes) has been set
up to probe more deeply for individual CO absorptions, at better signal to 
noise and at the full resolution of the SWS ($\sim$2500 around 4.7$\mu$m). 
Because of the way the SWS06 mode is defined, the 4.4-5.1$\mu$m range
was covered in repeated $\sim$40min long scans in between dark current 
measurements. These long scans are significantly affected by drifts and
jumps of the individual detector signals, at a level that can cause
broad structures at the few percent level in a standard processed average
spectrum. The standard processed average spectrum indeed shows weak broad
features at the $\sim$5\% level, at wavelengths not corresponding to the 
CO band or familiar solid state bands. We consider these structures
not reliable and specially processed the data in a way dealing with the
drift and jumps but maintaining information on possible individual 
absorption or emission lines. After standard dark current subtraction and
relative response calibration, jumps were identified in the time sequences
of the individual detectors and the pre- and post-jump signals smoothly 
connected by offsetting the post-jump data. We then subtracted from the flux 
history of each detector a median-smoothed flux history, with a smoothing width
corresponding to $\sim$0.03$\mu$m, and continued data reduction from then 
normally. The lower spectrum of Fig.~\ref{fig:sws} shows the resulting 
spectrum, offset by an arbitrary amount. We significantly detect the
[ArVI] 4.530$\mu$m line at the redshift of NGC\,1068, fainter non-significant
emission lines may be present at the positions of Pfund $\beta$ and of 
[NaVII] 4.685$\mu$m.

No significant individual absorption or emission lines from CO are
detected, to a 3$\sigma$ limit of 0.43Jy (7\% of the continuum) at the 
resolution 2500 of the SWS06 spectrum. We have probed for fainter 
band emissions or absorptions by crosscorrelating the NGC\,1068 spectrum 
(with the three emission lines masked out) with template CO `spectra' 
containing gaussians of appropriate width at the positions of CO 
band lines, up to a selectable maximum rotational level J. No 
clear positive or negative correlation peak near the systemic redshift was 
observed. This suggests that any clear band in emission or absorption must
be noticeably fainter than the limit for individual lines quoted above,
with exact limits a function of the adopted number of contributing 
rovibrational lines corresponding to the unknown physical conditions.

\section{Implications}
No clear signature of the CO 4.7$\mu$m band has been detected in the
spectra of bright local AGN. To our knowledge, no  full treatment of these 
lines has yet been published for one of the currently available `torus' models.
However, given the considerable uncertainties about the detailed structure
of this region and the significant continuum optical depth it is 
likely that configurations exist that would not yield absorptions or 
emissions that are
detected in our data. A dense and warm cirumnuclear region of AGN 
could show these lines in emission or absorption depending on the detailed
geometry and viewing angle, with contrast that is not necessarily high.

More significant, however, is the difference between the spectra
presented here and the spectrum of IRAS F00183-7111, itself most likely
hosting an obscured AGN (Spoon et al. \cite{spoon04}).
Our results suggest that the CO absorption seen in IRAS F00183-7111 is not
a straightforward signature of an AGN, but related to the
extraordinary concentration of dense and warm gas in the nuclear region of
ULIRGs. We speculate that the most heavily obscured (U)LIRGs also showing
strong continuous absorption features are most likely to show similar
behaviour - IRAS F00183-7111 is a prime example of this subclass among
the ULIRGs 
(Tran et al. \cite{tran01}, Spoon et al. \cite{spoon04}). Spoon et al.
(\cite{spoon04}) estimate from the CO absorption a colum of 
$\rm 10^{23.5}cm^{-2}$ of dense 
$\sim$700K molecular gas towards its nuclear region. Such gas will also 
emit in the mid-infrared rotational lines of molecular hydrogen. A realistic
model of this emission would assume a clumpy distribution of gas in which
our line of sight is not special with respect to the intercepted column 
density. However, for simplicity we have computed the emission from a 
geometrically and optically thin shell of dense gas of this 
column density surrounding the nuclear region. The measured flux
of the H$_2$ 0-0 S(3) line of IRAS F00183-7111 (Spoon et al. \cite{spoon04})
corresponds to $\rm\sim 10^5M_\odot$ of 700K gas at the distance of 
F00183-7111,
sufficient for a shell of radius $\sim$2pc shell with the column inferred
by Spoon et al. (\cite{spoon04}). This estimate
indicates that the obscuring warm/dense region is likely small, but
is subject to
the significant opposing uncertainties of (1) obscuration of the S(3) line
from this small component and (2) contributions to this line of other larger, 
and likely little obscured regions. A similar r$\sim$0.5pc scale is 
obtained assuming the 5$\mu$m continuum of IRAS F00183-7111 is optically thick
emission from 1000K dust, this is likely a lower limit to the size of the
CO absorbing region.

The challenge is to explain the 
dichotomy between spectral properties of IRAS F00183-7111 and nearby AGN: 
Some classical models for compact tori (e.g. Pier \& Krolik \cite{pier93}) 
invoke large ($\rm\sim 10^{24}cm^{-2}$) columns on parsec scales, which in 
Type 2 objects are seen in absorption against the nucleus. While there is still
uncertainty on the detailed structure and columns, the presence of warm 
dust on such scales is directly confirmed by mid-infrared interferometry 
of NGC\,1068 (Jaffe et al. \cite{jaffe04}), reinforcing earlier evidence
for dense parsec-scale circumnuclear gas on the basis of masers and 
radio continuum
emission (Gallimore et al. \cite{gallimore96}, \cite{gallimore97}). For 
both nearby AGN and IRAS F00183-7111,
large columns of warm gas on parsec scales are thus suggested but,
still, the observed spectral signatures differ strongly. For the nearby 
Seyferts, the 4-5$\mu$m emission may be dominated by relatively little obscured
warm dust, even if the column through the torus towards the very nucleus 
is high. For an axially symmetric `torus', such dust can be seen 
towards parts of the inner face of the axial opening of the `torus', even 
in Seyfert 2 objects, or from dust extended on larger scales in the 
narrow line region. If NGC 1068
is representative, such little obscured regions must not be strong CO 
emitters. It is important to
explore the role of geometric coverage in a realistic radiative transfer 
model including the CO lines. Models with coverage in all directions
may be the key to strong features like those seen in IRAS F00183-7111. 
A realistic modelling of an IRAS F00183-7111-like CO absorption around a
central obscured AGN/starburst could also provide better constraints on the 
size and structure of this region. Future Spitzer observations, in particular
of moderate redshift infrared galaxies and luminous AGN, will be able to 
shed more light on the differences to local AGN.

\begin{acknowledgements}
We thank Amiel Sternberg for discussions and the referee for helpful comments.
The ISO Spectrometer Data Center at MPE is supported by DLR under grant
50 QI 0202.

\end{acknowledgements}

\end{document}